# Ultrafast Thermal Modification of Strong Coupling in an Organic Microcavity

Bin Liu[1], Vinod M. Menon [1,2*], Matthew Y. Sfeir[2,3*]

[1]Department of Physics, City College of New York, New York, NY 10031, USA
[2]Department of Physics, Graduate Center, City University of New York, New York, NY 10016, USA
[3]Photonics Initiative, Advanced Science Research Center, City University of New York, New York, NY 10031, USA

* Email: vmenon@ccny.cuny.edu
* Email: msfeir@gc.cuny.edu

**Abstract:** There is growing interest in using strongly coupled organic microcavities to tune molecular dynamics, including the electronic and vibrational properties of molecules. However, very little attention has been paid to the utility of cavity polaritons as sensors for out-of-equilibrium phenomena, including thermal excitations. Here, we demonstrate that non-resonant infrared excitation of an organic microcavity system induces a transient response in the visible spectral range near the cavity polariton resonances. We show how these optical response can be understood in terms of ultrafast heating of electrons in the metal cavity mirror, which modifies the effective refractive index and subsequently the strong coupling conditions. The temporal dynamics of the microcavity are strictly determined by carriers in the metal, including the cooling of electrons via electron-phonon coupling and excitation of propagating coherent acoustic modes in the lattice. We rule out multiphoton excitation processes and verify that no real polariton population exists despite their strong transient features. These results suggest the promise of cavity polaritons as sensitive probes of non-equilibrium phenomena.

## INTRODUCTION

The strong coupling (SC) regime of light-matter interaction can be realized when an optically active material and a cavity resonator exchange energy reversibly before losses occur, giving rise to new hybrid bosonic quasiparticles called cavity polaritons. The cavity polariton is a superposition of the cavity photon and exciton state, and is characterized by the anti-crossing of two momentum-energy dispersion curves, referred to as the upper (UPB) and lower polariton branch (LBP) respectively.[1,2] Due to the large oscillator strength and binding energy of Frenkel excitons in organic materials, organic cavity polaritons can yield large vacuum Rabi splitting energies even at room temperature.[3–6] The hybrid light-matter character of organic cavity polaritons have made them attractive for observing diverse fascinating photonic phenomena such as polariton condensation and lasing,[7–9] optical harmonic generation enhancement,[10–12], superfluidity.[13] Beyond the dynamics of the polaritons themselves, much recent attention has focused on electronic interactions between organic cavity polaritons and the reservoir states (primarily exciton character), frequently with the goal of optimizing their excited state dynamics.[14–18] These include chemical processes for device applications, including polariton LEDs,[19,20] polariton-mediated energy transfer,[21–23] photochemical reaction rate and work function modification,[24,25] and transistor action.[26]

However, recent evidence has suggested that significant modification of the organic chemical dynamics may be difficult to achieve in practice. Pioneering experimental studies identified strong and persistent transient optical features corresponding to the cavity polariton resonances and proposed that the decay of these optical features directly reflected the polariton population dynamics, and the interpretation that the long observed lifetime is an intrinsic property of the hybrid state was widely adopted in the organic polariton community.[27–30] However, recent theoretical treatments have highlighted that photoexcitation at the LP optical transition still induces a significant population of reservoir states due to a large spectral overlap with polaritons and that these reservoir states will subsequently dominate the excited state relaxation dynamics.[31,32] Subsequent studies of organic cavity polaritons using fluorescence-based probes support the idea that exciton states dominate the relaxation dynamics, with long time scales for scattering of reservoir states to cavity polaritons on the order of > 100 ns.[33,34] Furthermore, we have recently demonstrated that the origin of the strong and persistent transient features of cavity polaritons that appear under resonant pumping actually arise from the population of *reservoir* states, which leads to transient modification of the strong coupling conditions.[35] In other words, the polariton lifetime cannot be directly determined from the decay of transient features corresponding to the cavity polaritons; instead, these optical transient response reflect the time-dependent changes in the effect refractive index due to a long-lived exciton population. Collectively, this recent evidence suggests that the cavity polariton modes function most effectively as sensitive probes of out-of-equilibrium phenomena, including transient exciton populations and their conversion dynamics.[27,35]

Here, we show that the strong coupling conditions in an organic microcavity can be transiently modified even under low photon energy *non-resonant* excitation conditions. Thermal excitation of the cavity mirrors with infrared light produces transient reflectivity features in the *visible* region of the spectrum, even when we use photon energies that are too small to excite the organic active layer. In agreement with our previous studies for resonant excitation, we show how transient optical features corresponding to cavity polariton transitions emerge due to changes in the strong coupling conditions, in this instance being caused by modification of the complex refractive index ($n$, $k$) of the top (semi-transparent) metal mirror. We use a simple two-temperature model to approximate the dynamical evolution of the electronic and lattice temperatures within the semi-transparent metal cavity mirror, and calculate the time-dependent complex refractive index within the Drude model. Using this picture, we reproduce the measured transient reflectivity spectra such that the fast early-time kinetics is caused by the rapid dissipation of electronic energy to the lattice, while the long-lived kinetics of cavity systems is governed by coherent vibrational motions of lattice phonons. The experimental observation of ultrafast thermal modification of the dynamics in strongly coupled organic microcavity, first reported here, provides a new means for real-time tuning of cavity polariton states, and may suggest potential photonic switching applications with external ultrafast thermal manipulation.

## EXPERIMENTAL METHODS

### Materials and Sample Fabrication

The organic molecule used is triisopropysilylacetylene pentacene (TIPS-Pc), with the molecular chemical structure shown in the inset of **Figure 1a**, as the excitonic material. TIPS-Pc has been widely studied in the context of organic optoelectronics, due to its relatively high carrier mobility and ability to undergo singlet exciton fission, a process in which a single absorbed photon is converted into two independent triplet excitons.[36] To obtain neat solid films, it is necessary to disperse it into a transparent polymeric matrix. A polymer matrix solution was first prepared using polystyrene (PS, Sigma-Aldrich, 182427-500G, Mw ~ 280,000) dissolved in toluene with a concentration of 15 mg/ml, and TIPS-Pc molecules were then added to the solution with a relative mass fraction of 30%, which could provide enough number of oscillators in the microcavity for realizing exciton-photon strong coupling.

For the all-metal microcavity fabrication, a 100 nm-thick Ag mirror was deposited by an e-beam evaporator in vacuum at $10^{-7}$ Torr, and the 160 nm TIPS-Pc /PS cavity layer was then spin-cast on top of the Ag mirror. Microcavity fabrication was completed by depositing a second 30 nm-thick Ag layer on top of the organic film. For the DBR-Ag microcavity fabrication, a DBR comprising 12 pairs of $Si_3N_4$ (80 nm)/$SiO_2$ (110 nm) was prepared by Plasma-enhanced chemical vapor deposition (PECVD), whose stop-band has a central wavelength of ~ 650 nm at normal incidence. The TIPS-Pc /PS layer was then spin-cast on top of DBR, and microcavity fabrication was completed by depositing a 30 nm-thick Ag layer on top of the organic film.

**Linear Optical Spectroscopy**

Strong coupling of the organic cavity was characterized using angle-resolved reflectivity and photoluminescence (PL) measurements at room temperature with a home-built Fourier space imaging setup.[37] For angle-resolved reflectivity measurements, a broadband of white light from tungsten halogen lamp was focused onto the sample by a 50x objective with a high numerical aperture (NA=0.8), and the reflected signal was collected by the same 50x objective, covering an angular range of ±53.1º, followed by the detection using the spectrometer (Princeton Instruments, Acton SpectraPro SP-2500) and charge-coupled device (CCD) camera (Princeton Instruments, PIX 1024B). For angle-resolved PL measurements, a fiber-coupled CW blue diode at 462 nm with a laser line filter was used to excite the sample, and the pump beam was focused onto the sample using the 50x objective (NA=0.8) with a spot size of 5 μm in diameter, and the emission signal was collected by the same 50x objective. A 600 nm long-pass filter was used to block the residual excitation beam.

**Transient Absorption Spectroscopy**

Transient absorption spectroscopy was performed using a commercial Ti:Sapphire laser system (800 nm, 1kHz repetition rate). A commercial optical parametric amplifier (TOPAS-C) was used to generate infrared pump pulses with approximately 80 fs pulse widths. For femtosecond pump-probe measurements, supercontinuum probe light was generated by focusing the 800 nm fundamental into a sapphire disc. The probe light was split into signal and reference beams, which were detected on a shot-by-shot basis by a fiber-coupled silicon visible diode array. The pump-probe delay was controlled by a mechanical delay stage. All transient absorption spectroscopy was performed using the reflection configuration with a small incident angle for the probe.

**RESULTS AND DISCUSSION**

**Figure 1a** shows the absorption and photoluminescence (PL) spectra measured from a TIPS-Pc /PS film spin-cast on a quartz substrate. We designed our cavity to achieve strong coupling with the lowest energy exciton state ($S_1$) near 650 nm (1.91 eV), resulting in an optimal film thickness of ∼ 160 nm that corresponds to a total optical density of ∼ 0.18 (**Figure 1a**). The photoluminescence of the neat film is typical of organic chromophores, with a small Stokes shift for the maximum at 655 nm and the characteristic set of vibronic peaks forming a mirror image of the absorption (**Figure 1a**). **Figure 1b** shows structures for two types of microcavities, the all-metal cavity and the dielectric-metal cavity, which are typical Fabry-Perot resonators. The geometry of these two cavities allows optical probing through the semi-transparent top reflector. We have recently examined the transient optical response of an equivalent all-metal of TIPS-Pc microcavity under resonant pumping conditions, concluding that the cavity polaritons have negligible effect on the singlet fission dynamics of TIPS-Pc.[35]

**Figure 2a** shows the contour map of the TE-polarized angle-resolved reflectivity for the all-metal (left) and DBR-Ag cavity (right), respectively, revealing the anti-crossing dispersion of upper (UP, cyan) and lower polaritons (LP, red) in the exciton-photon strong coupling regime**.** The same dispersion is found in the TE-polarized angle-resolved photoluminescence data from both the all-metal and DBR-Ag cavities (**Figure 2d** and **2e**), indicating the dominant radiative relaxation channel occurs through the LP. The polariton dispersion for the UP and LP as a function of angle, $E_{UP,LP}(\theta)$, was determined using a coupled harmonic oscillator model given by[2]

$$\begin{pmatrix} E_C(\theta) - i\gamma_C & V \\ V & E_{Ex} - i\gamma_{Ex} \end{pmatrix} \begin{pmatrix} \alpha \\ \beta \end{pmatrix} = E_{UP,LP}(\theta) \begin{pmatrix} \alpha \\ \beta \end{pmatrix} \qquad (1)$$

$$E_{UP,LP}(\theta) = \frac{1}{2} \left[ E_C(\theta) + E_{Ex} - i(\gamma_C + \gamma_{Ex}) \right] \pm \sqrt{V^2 + \frac{1}{4}[E_C(\theta) - E_{Ex} + i(\gamma_C - \gamma_{Ex})]^2} \qquad (2)$$

Here, the cavity mode energy ($E_C$) is approximated by $E_C(\theta) = E_0(1-\sin^2\theta/n^2_{\text{eff}})^{-1/2}$, where $E_0$ is the cavity cut-off energy (at normal), the uncoupled exciton energy ($E_{Ex}$) is fixed by the $S_1$ energy of TIPS-Pc at 1.91 eV, and the linewidth of cavity mode and molecular exciton ($\gamma_C$ and $\gamma_{Ex}$, respectively) are extracted from the experimental reflections and absorption curves. The coupling strength of cavity photons to excitons ($V$) and the Hopfield coefficients $|\alpha|^2$ and $|\beta|^2$ (**Figure 2b** and **2c**) that represent the fraction contribution of cavity photon and exciton component are extracted using a global fit of the UP and LP reflectivity minima. This procedure yields values for the Rabi splitting, $\hbar\Omega_R = 110 \pm 10$ meV for the all-metal and $100 \pm 10$ meV for the DBR-Ag cavity. As DBR-Ag cavity has a larger cavity quality factor ($Q$) and smaller linewidth compared to all-metal cavity, the Rabi splitting from DBR-Ag cavity is slightly smaller due to the greater linewidth mismatch (**Equation 2**) between cavity mode and molecular excitons. Two additional small differences between the two cavities are apparent in the reflectivity spectra, but do not affect the excited state dynamics discussed subsequently. In the all-metal cavity, additional strong coupling interactions occur between the vibronic overtone of the singlet exciton at 600 nm and cavity photons at large angles (> 40°, left panel in **Figure 2a**). Furthermore, modes from the DBR sidebands are apparent as dispersive reflectivity features in the spectral region of 500 - 550 nm and > 700 nm (**Figure 2a,** right).

To quantify differences in the strong coupling between the equilibrium (without optical pump) and non-equilibrium (with optical pump) states of the organic cavity, we employ transient reflectivity spectroscopy using ultrafast pump-probe techniques with low-energy, non-resonant pump pulses.[38] For these measurements, a tunable infrared pump beam is incident on the sample through the semi-transparent top silver mirror, and the visible broadband probe beam is collected in a reflective geometry. The pump-modulated reflectivity is plotted as $-(\Delta R/R)$, with positive quantities corresponding to transiently suppressed reflectivity and negative quantities corresponding to transiently enhanced reflectivity. To calibrate the incident probe angle and quantify the strong exciton-photon coupling in the pump-probe setup, we first measure the reflectivity spectrum of the all-metal cavity with the pump beam blocked. As in the Fourier-space measurements, we observe two strong reflectivity dips (inset of **Figure 3a**) indicating the upper (630 nm) and lower polariton (665 nm) state, respectively. These values for the LP and UP correspond to an incident angle of $\sim 10°$.

The transient spectra and dynamical evolution of the system under infrared excitation are remarkably different from the results obtained with the pump photon energy tuned to excite the exciton or polariton states.[35] For example, for a pump pulse wavelength of 1340 nm (0.93 eV) incident on the microcavity through the semi-transparent top Ag mirror, the photon energy is too low to excite the LP. Still, measurable signals are observed in the visible spectral region, primarily as sharp derivative-like features localized near the UP (630 nm) and LP (665 nm) optical transitions (**Figure 3a**). Importantly, there are no signatures of the molecular excitons at any time. Particularly notable for its absence is the strong and long-lived (> 1 μs) triplet-triplet absorption feature at $\sim 510$ nm that results from singlet exciton fission in this material, which we have observed for resonant pumping of either the cavity polariton or reservoir exciton states (**Figure**

**3b**). Similarly, the kinetics exhibit a strong fast decay containing most of the initial amplitude, followed by a weak, slow decay extending for > 100 ps. This is in strong contrast to resonant pumping, where distinct singlet and triplet exciton dynamics are observed and which evolve over many orders of magnitude in time, from pico- to microseconds. We observe nearly identical transient behavior in the all-metal cavity and in the DBR-Ag cavity (**Figure 4a**) under the same excitation conditions. Additional features in the transient reflectivity spectra of the DBR-Ag cavity, for example near 530 nm and 565 nm, correspond to the DBR sidebands (**Figure 2a**, right panel) rather than any underlying excitation of the organic film.

Besides the lack of transient features identified with an exciton population and the distinct carrier dynamics observed under resonant pumping, there are several additional pieces of evidence that support a distinct excitation mechanism for the microcavity. First, we note that we need to use infrared pump fluences that are 10-100× larger than resonant excitation conditions to obtain comparable signal levels. Furthermore, the intensity of the signals near the cavity polaritons is non-linear with the incident fluence, unlike typical transient absorption experiments on molecular systems (**Figure 4b**). We rule out two-photon excitation, which would also give a non-linear transient response, by tuning the photon to well below the two-photon absorption onset (**Figure 4c**), which for TIPS-Pc occurs at ~ 1200 nm.[39,40] The system dynamics response is fully independent of the infrared pump wavelength, which we tune from 1250 nm (0.99 eV) – 1500 nm (0.83 eV), and instead, find that the transient signals depend only on the incident fluence level (**Figure 4d**).

Together, this evidence suggests that the physical origin of the transient features observed under low-energy infrared pumping arises from the impulsive thermal excitation of electrons in the semi-transparent top metal mirror. This thermal excitation mechanism would account for the wavelength independent, non-linear signal levels observed in the data, and induce transient signals near the LP and UP transitions *via* modifications to the refractive index ($n$, $k$) of the top Ag mirror by the infrared pump pulse. The carrier dynamics of simple metals, such as the semi-transparent top Ag mirror in our devices, are well approximated by the two-temperature model (TTM),[41–44] in which the absorption of a photon results in a hot thermalized electron distribution that subsequently equilibrates to the lattice via electron-phonon coupling. In the TTM model, the metal is divided into two subsystems: conduction electrons that are thermalized by (Coulomb) electron-electron interactions characterized by a temperature $T_e$, and the ionic lattice with a characteristic temperature $T_l$ that is thermalized by (anharmonic) phonon-phonon interactions.[44] The flow of heat between electrons and lattice occurs when $T_e > T_l$, with a time evolution given by two coupled differential equations:[41–44]

$$\gamma T_e \frac{\partial T_e}{\partial t} = -g(T_e - T_l) + S(t)$$

$$C_l \frac{\partial T_l}{\partial t} = g(T_e - T_l) \tag{3}$$

Here, $\gamma$ is the electron heat capacity constant (65 J m$^{-3}$ K$^{-2}$ for Ag)[42] and $C_l$ is the phonon heat capacity (2.43 $\times$ 10$^6$ J m$^{-3}$ K$^{-1}$). The source term describes the energy deposition by the laser pulse and is given by $S(t) = J \cdot A \cdot f(t)/d$, where $J$ is the laser fluence, $A$ is the absorbance, $d$ is the thickness of Ag film, and $f(t) = \frac{2\sqrt{ln2}}{\sqrt{\pi}t_p} exp\left[-4ln2\left(\frac{t}{t_p}\right)^2\right]$ is the temporal laser pulse profile with the pulse width $t_p$. The electron-phonon coupling constant $g$ is treated as a fitting parameter in this model. Within the Drude model, we can calculate the time-dependent damping rate, $\Gamma_0 = A_{ee}T_e^2 + B_{ep}T_l$, (inset of **Figure 3c**) of Ag after photoexcitation as a function of the electronic and lattice temperatures, using previously determined $A_{ee}$ and $B_{ep}$ scattering coefficients.[41] This is followed by the computation of the full time-dependent complex refractive index ($n'$, $k'$) using the Drude model for intraband transitions and accounting for contributions from interband transitions according to the methodology reported by Rakić et al.[45] The reflectivity spectra of the all-metal cavity system is simulated by the transfer matrix method,[46,47] using the measured refractive index of organic thin film together with the calculated modified ($n'$, $k'$) of Ag at different times. Finally, the modified reflectivity is used to calculate the simulated differential transient reflectivity as -$\Delta R/R$.

The TTM model allows us to calculate the time-dependent changes in the reflectivity spectrum as a function of the incident infrared pump fluence and time. For example, for the incident laser fluence used to acquire the data in **Figure 3a,b** (500 µJ/cm$^2$), the electronic temperature impulsively rises by ~ 110 K, followed by rapid thermalization to elevate the lattice temperature by ~ 1 K in (**Figure 3c**). The corresponding change in the Drude scattering rate ranges from ~ 12% at early times (5.5 meV) to ~ 2% after the lattice heats up (~ 1 meV). Under these conditions, the simulated differential transient spectra from the all-metal cavity (**Figure 3d**) are a good match to the measured experimental spectra (**Figure 3b)** at different times. The dominant spectral dynamics are captured within this simple model, including the overall magnitudes of the signal (~ 0.2% maximum change in the reflectivity) and their decay dynamics, in which the majority of the transient reflectivity signal decays in ~ 1.5 ps followed by a weaker, slow decay component. We note that the value for the electron-phonon coupling factor ($g = 3.4 \times 10^4$ J ps$^{-1}$ K$^{-1}$ m$^{-3}$) that reproduces the observed experimental kinetic evolution is similar to those reported for other Ag nanostructures at these fluences.[48] We note that this simple model also captures the primary spectral dynamics, including the weak broad absorption in the region of 450 – 550 nm and 700 – 750 nm, and the bleaches near the UP and LP features.

The transient absorption spectra of the DBR-Ag cavity excited by an infrared pump pulse can be understood using the same framework as the all-metal cavity, with excitation of the top semi-transparent Ag mirror leading to rapidly decaying spectral features corresponding to the LP transition at ~ 670 nm (**Figure 4a**). We note that due to different detuning, the UP is not prominent in the reflectivity spectrum (**Figure 2a** (Right) and inset of **Figure 4a**) at this particular angle (~ 15°). Here, we see that the peak intensity of transient spectra near lower polariton wavelength nonlinearly increases with increasing pump fluence, with the amplitude depending only on the incident laser fluence and not on pump wavelength. For example, the transient spectra are identical for the same fluence (ranging from 125 – 625 µJ/cm$^2$) at the same delay time (0.1 ps) for a pump wavelength of either 1340 nm (**Figure 4b**) or 1500 nm (**Figure 4c**), respectively. The pump-wavelength independent scaling behavior is summarized in **Figure 4d**, for 1250 nm, 1340 nm, and 1500 nm pump wavelengths. We note that the observed nonlinearity scaling of $\Delta R/R$ could result from either the nonlinear scaling of the temperature with input fluence or the complex dependence of the refractive index on the electronic and lattice temperatures.[41] It is worth mentioning that this ultrafast thermal effect is negligible under resonant pumping conditions with visible pump wavelengths at the input fluences (~ 25 µJ/cm$^2$) used in previous studies,[35] since the electronic temperature of the Ag mirror rises minimally (ΔT ~ 5 K) and the lattice temperature is essentially unchanged (ΔT < 0.1 K).

In addition to the fast electron dynamics in the first 1-2 ps, another characteristic of the carrier dynamics of thin metal films and nanostructures is the presence of coherent phonon oscillations on longer time scales. Here, we find that signatures of these coherent phonons are also present near the cavity polariton resonances for non-resonant infrared pumping, emerging as regular oscillations in our transient reflectivity data (**Figure 5a**). For 1340 nm pumping at 500 µJ/cm2 (same excitation conditions as **Figure 3a**), a time-dependent oscillation with period of ~ 20 ps is unambiguously observed near both UP and LP state (**Figure 5b**), similar to what is observed in previous studies on Ag film and nanoparticles and assigned to coherent phonon oscillations.[48–50] This observation verifies that, similar to the short-lived kinetic feature, the long-lived dynamics of strongly coupled organic cavity system with a semi-transparent Ag mirror under non-resonant infrared pumping conditions is also determined by the ultrafast thermal modification of the metal mirror.

**CONCLUSIONS**

We experimentally demonstrate the effect of low-energy, non-resonant pumping on the exciton-photon coupling in organic all-metal cavity and DBR-Ag cavities at room temperature. We observe that the infrared optical pumping is sufficient to modify the refractive index of the metal mirror of organic cavities, which transiently modifies the strong exciton-photon coupling conditions, and leads to the optical transient response near the hybrid polariton states. Calculations using a simple two-temperature model imply that the dynamical evolution of electronic and lattice temperature of the metal mirror govern the short- and long-lived kinetic behavior of cavity systems, respectively, and agree well with changes in the time-resolved reflectivity. In contrast to what is usually considered, this study shows that changes in the effective refractive index $n_{eff}$ of the cavity system can result from energy storage anywhere within the device, including thermal excitation of the mirrors. The observation of ultrafast thermal modification of the dynamics in strongly coupled cavity may suggest potential photonic switching applications with external ultrafast thermal manipulation.

**Corresponding Author**

* Email: vmenon@ccny.cuny.edu, msfeir@gc.cuny.edu.

**Acknowledgements**

The authors acknowledge support from the U.S. Department of Energy, Office of Basic Energy Sciences through Award No. DE-SC0017760. This research used resources of the Center for Functional Nanomaterials, which is a U.S. DOE Office of Science Facility, at Brookhaven National Laboratory under Contract No. DE-SC0012704. The authors also acknowledge the use of the Nanofabrication Facility at the Advanced Science Research Center of the City University of New York for the device fabrication.

**Data Availability Statement**

The data that support the findings of this study are available from the corresponding author upon reasonable request.

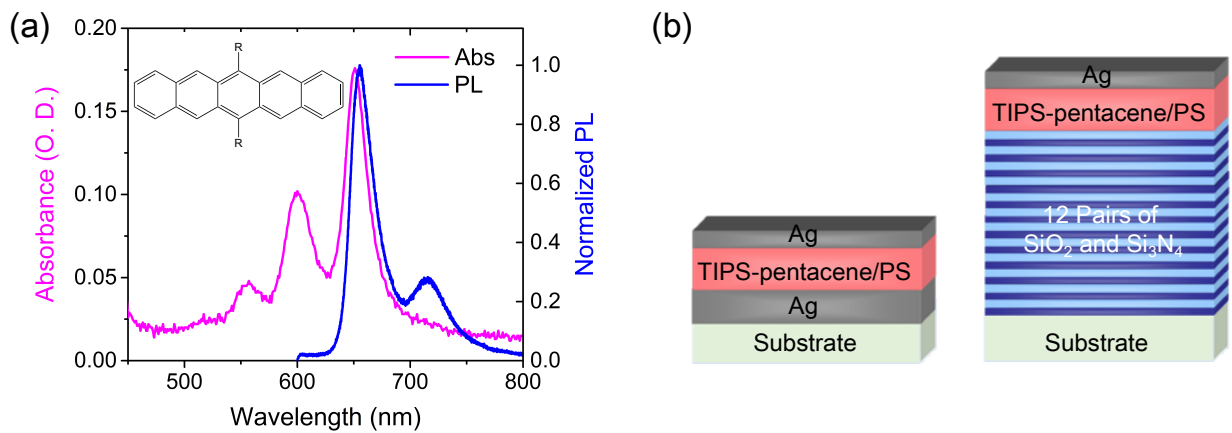

**Figure 1**. a) The absorption (magenta) and photoluminescence spectrum (blue) of a 160 nm thick film of TIPS-Pc molecules dispersed into a polystyrene matrix. Inset: the chemical structure of TIPS-Pc. b) Microcavity structures: All-metal cavity (left) and DBR-Ag cavity (right).

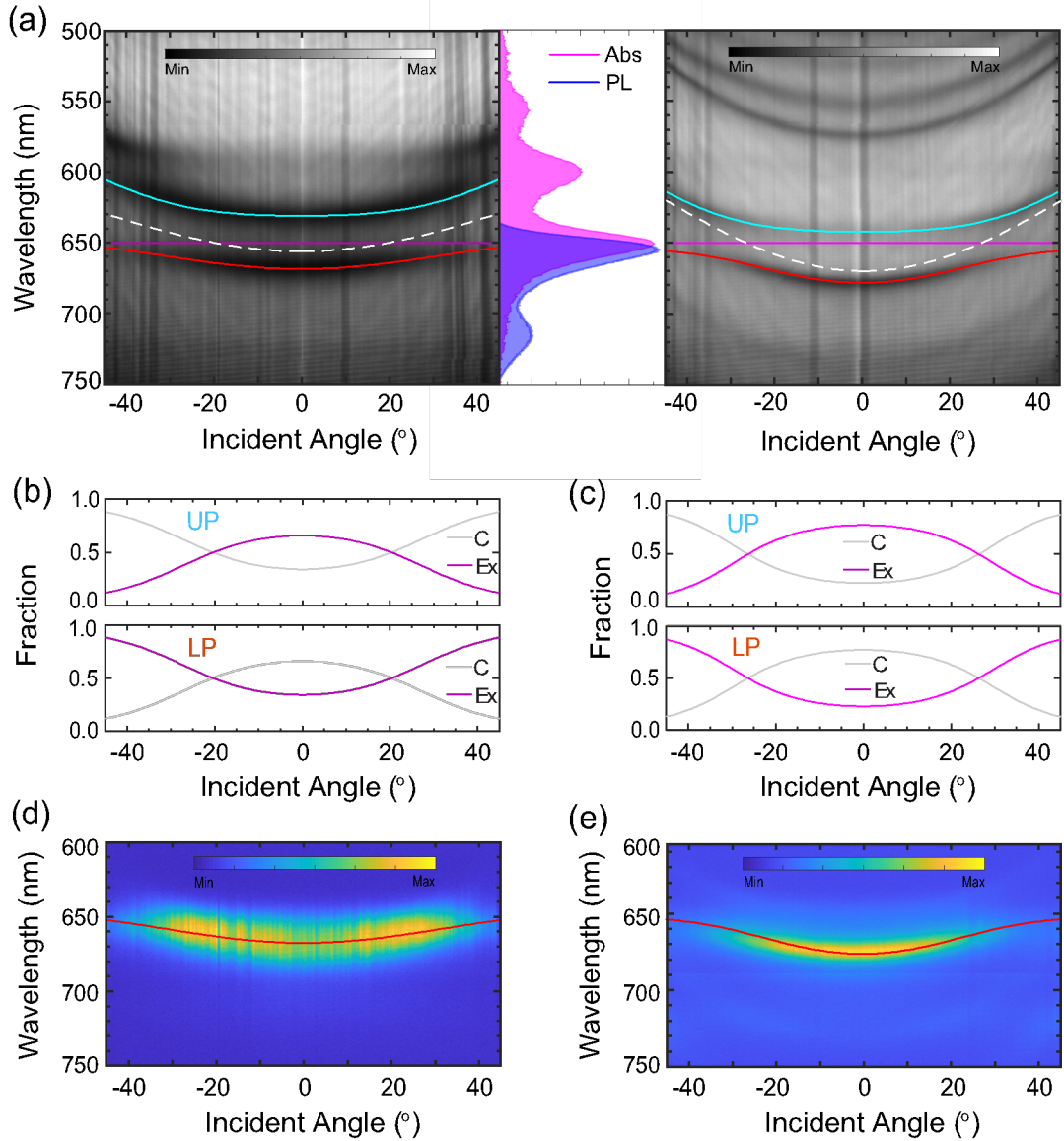

**Figure 2**. a) TE-polarized angle-resolved reflectivity map from an all-metal cavity (Left), and a DBR-Ag cavity (Right). The solid magenta and white curves show the uncoupled exciton and cavity photon dispersion; the solid cyan and red curves trace the calculated polariton dispersion using a coupled oscillator model. Middle: The absorption (magenta) and PL spectrum (blue) of a 160 nm TIPS-Pc/PS film. b) Top: Hopfield coefficients $|\alpha|^2$ and $|\beta|^2$ showing the compositions of UP and LP from the all-metal cavity, where the gray and purple curve represent the fraction of cavity photon and exciton component, respectively. c) Hopfield coefficients $|\alpha|^2$ and $|\beta|^2$ from the DBR-Ag cavity. d) TE-polarized angle-resolved photoluminescence map of all-metal cavity, and the solid red curve indicate the lower polariton dispersion. e) TE-polarized angle-resolved photoluminescence map of DBR-Ag cavity.

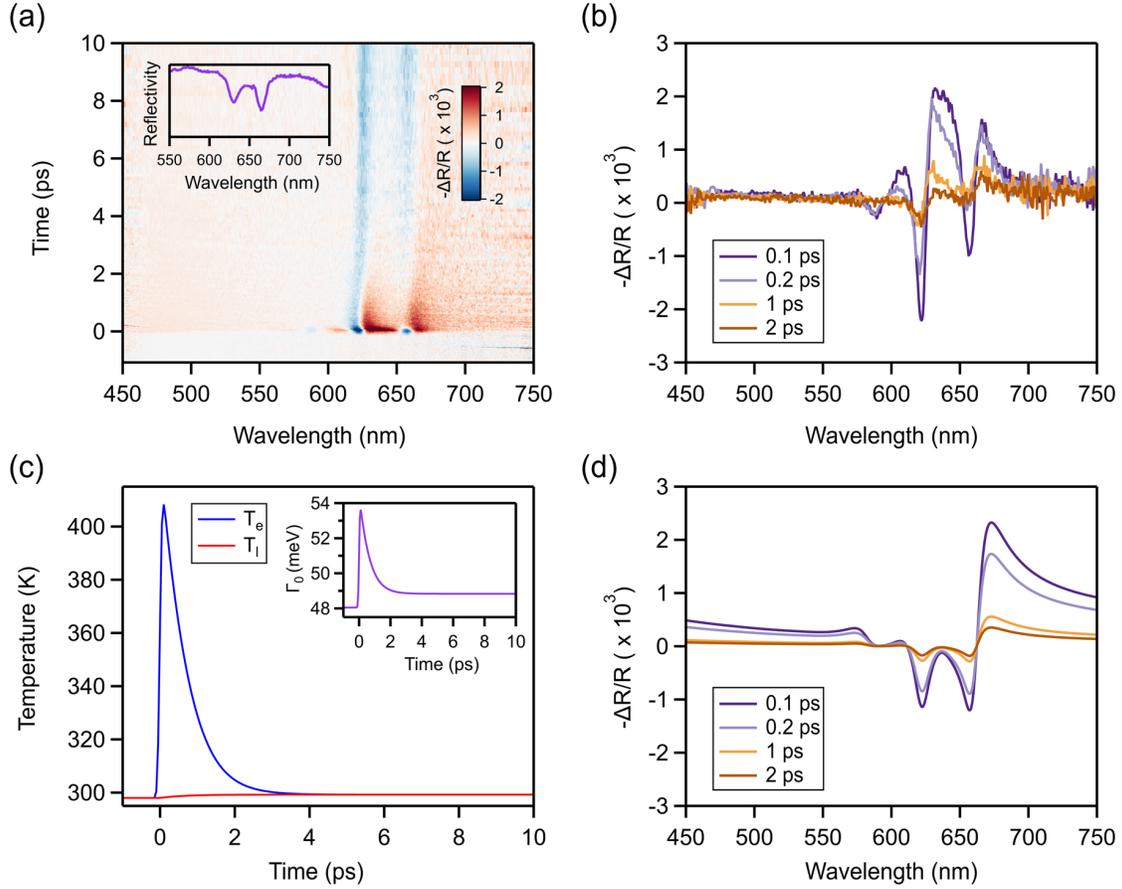

**Figure 3**. a) Transient spectra data of strongly coupled all-metal cavity with singlet fission molecules, excited with the wavelength of 1340 nm and fluence of 500 μJ /cm². Inset: reflectivity spectrum of the all-metal cavity measured only using the probe beam at 10° in the pump-probe setup. b) Transient spectra extracted from the raw data set at different times between 0.1 and 2 ps. c) Time-dependent electronic and lattice temperature of Ag from two-temperature model, with the electron-phonon coupling constant $g = 3.4 \times 10^4$ J ps⁻¹ K⁻¹ m⁻³. Inset: time-dependent broadening (inverse scattering time) calculated using the Drude model. d) Calculated transient spectra at different times between 0.1 and 2 ps using transfer matrix method with modified refractive index of Ag determined from the Drude model with the temperature evolution in c).

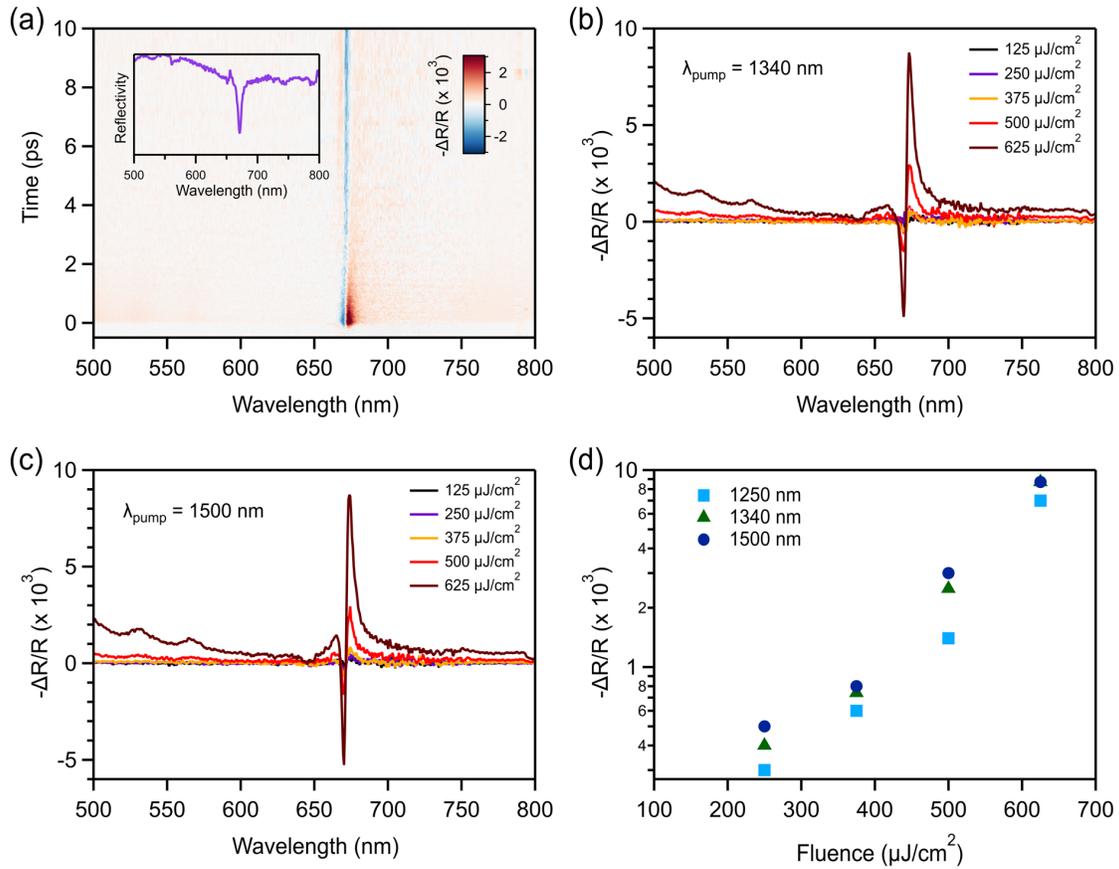

**Figure 4**. a) Transient spectra data of strongly coupled DBR-Ag cavity with singlet fission molecules, excited with the wavelength of 1340 nm and fluence of 500 μJ /cm². Inset: reflectivity spectrum of the DBR-Ag cavity measured only using the probe beam at 15° in the pump-probe setup. b) Fluence-dependent transient spectra at the time of 0.1 ps with the infrared pump wavelength of 1340nm, and (c) 1500 nm. d) Experimental transient spectral peak intensity versus excitation fluence for three different infrared pump wavelengths.

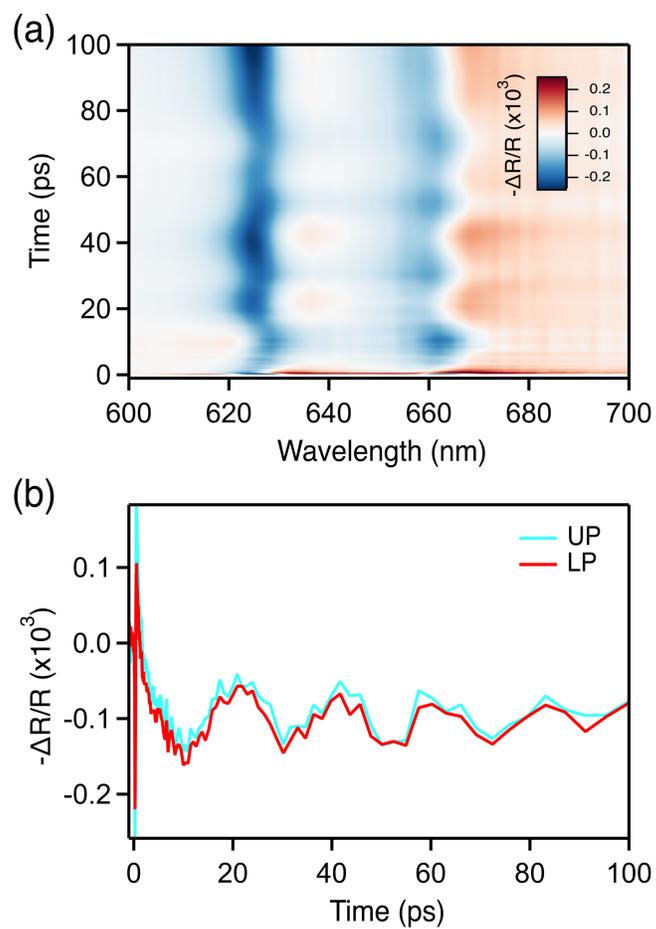

**Figure 5**. a) Transient spectra data obtained from the same measurement as shown in **Figure 3a** but with longer time-scale results. b) Comparison of kinetic slices near UP and LP wavelength.

Reference:

[1] C. Weisbuch, M. Nishioka, A. Ishikawa, and Y. Arakawa, Phys. Rev. Lett. **69**, 3314 (1992).

[2] M. S. Skolnick, T. A. Fisher, and D. M. Whittaker, Semicond. Sci. Technol. **13**, 645 (1998).

[3] D. G. Lidzey, D. D. C. Bradley, M. S. Skolnick, T. Virgili, S. Walker, and D. M. Whittaker, Nature **395**, 53 (1998).

[4] R. J. Holmes and S. R. Forrest, Org. Electron. Physics, Mater. Appl. **8**, 77 (2007).

[5] S. Kéna-Cohen, S. A. Maier, and D. D. C. Bradley, Adv. Opt. Mater. **1**, 827 (2013).

[6] B. Liu, P. Rai, J. Grezmak, R. J. Twieg, and K. D. Singer, Phys. Rev. B **92**, 155301 (2015).

[7] J. D. Plumhof1, T. Stöferle, L. Mai, U. Scherf, and R. F. Mahrt, Nat. Mater. **13**, 247 (2014).

[8] K. S. Daskalakis, S. A. Maier, R. Murray, and S. Kéna-Cohen, Nat. Mater. **13**, 271 (2014).

[9] T. Cookson, K. Georgiou, A. Zasedatelev, R. T. Grant, T. Virgili, M. Cavazzini, F. Galeotti, C. Clark, N. G. Berloff, D. G. Lidzey, and P. G. Lagoudakis, Adv. Opt. Mater. **5**, 1 (2017).

[10] T. Chervy, J. Xu, Y. Duan, C. Wang, L. Mager, M. Frerejean, J. A.W. Münninghoff, P. Tinnemans, J. A. Hutchison, C. Genet, A. E. Rowan, T. Rasing, and T. W. Ebbesen, Nano Lett. **16**, 7352 (2016).

[11] F. Barachati, J. Simon, Y. A. Getmanenko, S. Barlow, S. R. Marder, and S. Kéna-Cohen, ACS Photonics **5**, 119 (2018).

[12] B. Liu, M. Crescimanno, R. J. Twieg, and K. D. Singer, Adv. Opt. Mater. **7**, 1 (2019).

[13] G. Lerario, A. Fieramosca, F. Barachati, D. Ballarini, K. S. Daskalakis, L. Dominici, M. De Giorgi, S. A. Maier, G. Gigli, S. Kéna-Cohen, and D. Sanvitto, Nat. Phys. **13**, 837 (2017).

[14] F. Herrera and F. C. Spano, Phys. Rev. Lett. **116**, 238301 (2016).

[15] R. F. Ribeiro, L. A. Martínez-Martínez, M. Du, J. Campos-Gonzalez-Angulo, and J. Yuen-Zhou, Chem. Sci. **9**, 6325 (2018).

[16] J. Feist, J. Galego, and F. J. Garcia-Vidal, ACS Photonics **5**, 205 (2018).

[17] D. Sanvitto and S. Kéna-Cohen, Nat. Mater. **15**, 1061 (2016).

[18] T. W. Ebbesen, Acc. Chem. Res. **49**, 2403 (2016).

[19] M. Mazzeo, A. Genco, S. Gambino, D. Ballarini, F. Mangione, O. Di Stefano, and G. Gigli, Appl. Phys. Lett. **104**, 233303 (2014).

[20] C. R. Gubbin, S. A. Maier, and S. Kéna-Cohen, Appl. Phys. Lett. **104**, 233302 (2014).

[21] D. M. Coles, N. Somaschi, P. Michetti, C. Clark, P. G. Lagoudakis, P. G. Savvidis, and D. G. Lidzey, Nat. Mater. **13**, 712 (2014).

[22] K. Georgiou, P. Michetti, L. Gai, M. Cavazzini, Z. Shen, and D. G. Lidzey, ACS Photonics **5**, 258 (2017).

[23] X. Zhong, T. Chervy, L. Zhang, A. Thomas, J. George, C. Genet, J. A. Hutchison, and T. W. Ebbesen, Angew. Chemie - Int. Ed. **56**, 9034 (2017).

[24] J. A. Hutchison, A. Liscio, T. Schwartz, A. Canaguier-Durand, C. Genet, V. Palermo, P. Samorì, and T. W. Ebbesen, Adv. Mater. **25**, 2481 (2013).

[25] A. Canaguier-Durand, E. Devaux, J. George, Y. Pang, J. A. Hutchison, T. Schwartz, C. Genet, N. Wilhelms, J.-M. Lehn, and T. W. Ebbesen, Angew. Chemie Int. Ed. **52**, 10533 (2013).

[26] A. V. Zasedatelev, A. V. Baranikov, D. Urbonas, F. Scafirimuto, U. Scherf, T. Stöferle, R.F. Mahrt, and P.G. Lagoudakis, Nat. Photonics 13, 378 (2019).

[27] T. Schwartz, J. A. Hutchison, J. Léonard, C. Genet, S. Haacke, and T. W. Ebbesen, ChemPhysChem **14**, 125 (2013).

[28] J. George, S. Wang, T. Chervy, A. Canaguier-Durand, G. Schaeffer, J. M. Lehn, J. A. Hutchison, C. Genet, and T. W. Ebbesen, Faraday Discuss. **178**, 281 (2015).

[29] H. Wang, H. Y. Wang, A. Bozzola, A. Toma, S. Panaro, W. Raja, A. Alabastri, L. Wang, Q.

D. Chen, H.L. Xu, F. De Angelis, H. B. Sun, and R. P. Zaccaria, Adv. Funct. Mater. **26**, 6198 (2016).

[30] C. A. Delpo, B. Kudisch, K. H. Park, S. U. Z. Khan, F. Fassioli, D. Fausti, B. P. Rand, and G. D. Scholes, J. Phys. Chem. Lett. **11**, 2667 (2020).

[31] J. del Pino, F. A. Y. N. Schröder, A. W. Chin, J. Feist, and F. J. Garcia-Vidal, Phys. Rev. Lett. **121**, 227401 (2018).

[32] G. Groenhof, C. Climent, J. Feist, D. Morozov, and J. J. Toppari, J. Phys. Chem. Lett. **10**, 5476 (2019).

[33] D. Polak, R. Jayaprakash, T. P. Lyons, L. A. Martínez-Martínez, A. Leventis, K. J. Fallon, H. Coulthard, D. G. Bossanyi, K. Georgiou, A. J. Petty, J. Anthony, H. Bronstein, J. Yuen-Zhou, A. I. Tartakovskii, J. Clark, and A. J. Musser, Chem. Sci. **11**, 343 (2020).

[34] E. Eizner, L. A. Martínez-Martínez, J. Yuen-Zhou, and S. Kéna-Cohen, Sci. Adv. **5**, eaax4482 (2019).

[35] B. Liu, V. M. Menon, and M. Y. Sfeir, ACS Photonics 7, 2292 (2020).

[36] M. B. Smith and J. Michl, Chem. Rev. **110**, 6891 (2010).

[37] B. Liu, R. Wu, and V. M. Menon, J. Phys. Chem. C **123**, 26509 (2019).

[38] S. N. Sanders, E. Kumarasamy, A. B. Pun, M. T. Trinh, B. Choi, J. Xia, E. J. Taffet, J. Z. Low, J. R. Miller, X. Roy, X. Y. Zhu, M. L. Steigerwald, M. Y. Sfeir, and L. M. Campos, J. Am. Chem. Soc. **137**, 8965 (2015).

[39] E. Garoni, J. Zirzlmeier, B. S. Basel, C. Hetzer, K. Kamada, D. M. Guldi, and R. R. Tykwinski, J. Am. Chem. Soc. **139**, 14017 (2017).

[40] K. Kamada, K. Ohta, T. Kubo, A. Shimizu, Y. Morita, K. Nakasuji, R. Kishi, S. Ohta, S. I. Furukawa, H. Takahashi, and M. Nakano, Angew. Chemie - Int. Ed. **46**, 3544 (2007).

[41] A. N. Smith and P. M. Norris, Appl. Phys. Lett. **78**, 1240 (2001).

[42] R. H. M. Groeneveld, R. Sprik, and A. Lagendijk, Phys. Rev. B **51**, 11433 (1995).

[43] R. H. M. Groeneveld, R. Sprik, and A. Lagendijk, Phys. Rev. B **45**, 5079 (1992).

[44] S. I. Anisimov, B. L. Kapeliovich, and T. L. Perel'Man, J. Exp. Theor. Phys. **39**, 375 (1974).

[45] A. D. Rakić, A. B. Djurišić, J. M. Elazar, and M. L. Majewski, Appl. Opt. **37**, 5271 (1998).

[46] L. A. A. Pettersson, L. S. Roman, and O. Inganäs, J. Appl. Phys. **86**, 487 (1999).

[47] B. Liu, C. M. M. Soe, C. C. Stoumpos, W. Nie, H. Tsai, K. Lim, A. D. Mohite, M. G. Kanatzidis, T. J. Marks, and K. D. Singer, Sol. RRL **1**, 1700062 (2017).

[48] C. Voisin, N. Del Fatti, D. Christofilos, and F. Vallée, J. Phys. Chem. B **105**, 2264 (2001).

[49] M. van Exter and A. Lagendijk, Phys. Rev. Lett. **60**, 49 (1988).

[50] W. Qian, H. Yan, J. J. Wang, Y. H. Zou, L. Lin, and J. L. Wu, Appl. Phys. Lett. **74**, 1806 (1999).